\documentstyle[12pt]{article}

\begin{document} 

\begin{center}
{\bf \Large Complex temperatures zeroes of partition function in
spin-glass models.}\\
\vspace{4mm}
D.B. Saakian$^{1,2)}$, E.E. Vogel$^{2)}$.\\

$^{1)}$Yerevan Physics Institute,
Alikhanian Brothers St. 2,\\ Yerevan 375036, Armenia \\
$^{2)}$Universidad de La Frontera, Departamento de Cienas Fisicas,
\\Casilla
54-D, Temuco, Chile.
\end{center}

\begin{abstract}
An approximate method is proposed for investigating complex-temperature
properties of real-dimensional spin-glass models.
The method uses the complex-temperature data of the ferromagnetic model on
the same lattice.
The universality line in the complex-temperature space is obtained.

\end{abstract}

\section{Introduction.}

Investigating of physical quantities in the complex plane can reveal new
and unexpected effects.
More than four decades have passed since Dyson \cite{dyson}  
considered imaginary charges in electrodynamics.
Having $\alpha=e^2/ \hbar c < 0$, he got that this theory should be
essentially unstable,
and expansions in $\alpha$ are,
at best, asymptotic. The consideration of complex magnetic field by Lee
and Yang and complex temperatures
by Fisher \cite{fisher} opened new and effective method for investigating
phase transitions.
Later the method found large variety of applications in obtaining the
order and type of 
possible phase transitions \cite{lee}, critical indices
\cite{derrida3}\cite{zuber}, and recently has been
connected with experiments on magnetization \cite{binek}.

The method can be even more useful when investigating phase transitions in
disordered systems, since many well-developed 
analytical and numerical methods are not suitable here. This point of view
was first provided in \cite{grif} 
where so-called Graffiths' singularities
was discovered in statistical systems with random-fields. Recently, a
strongly-frustrated but nonrandom system was investigated in complex
temperature plane \cite{allah2}, and a continuous spectra of phase
transitions was obtained.

The consideration of spin-glasses in complex-field and/or
complex-temperature plane was started in \cite{derrida2}\cite{parga}.
Namely, Random Energy Model (REM) was investigated, which is the simplest
but typical representative of spin-glasses \cite{derrida1}.
This has been continued in \cite{david1} where the more physical dilute
(finite-connectivity) REM 
\cite{dom}\cite{allah1} was investigated. In particular, we have pointed
out also how the Dilute Generallized Random Energy Model (DGREM) should be
considered in the complex planes
\cite{david1}. This seems very important, because DGREM 
is nothing else, but the quite accurate approximation to the
real-dimensional (non-mean-field) Edwards-Anderson (EA) spin-glass model.
We believe that the results obtained with DGREM can be relevant for
that model, and will help to clarify the properties of spin-glasses
in finite dimensions. Recently, the lower critical dimension of the EA
model has been estimated in this fashion  \cite{david2}.
This line of research will be continued in the present paper. We shall
give the density of the partition function's zeroes, and 
discuss in details its applications to the real-dimensional EA models.

\section{Dilute REM }

We shall start with repeating some facts about diluted REM and GREM.

The dilute p-spin glass model is described by the following hamiltonian
\begin{equation}
\label{e1}
H=-\sum_{ (i_1<i_2..<i_p)=1}^{\alpha N}j_{i_1..i_p}s_{i_1}..s_{i_p},
\end{equation}
where only $\alpha N$ 
couplings $j_{i_1..i_p}$
are non-zero $1\geq i \geq N$, and $s_i=\pm 1$ are Ising spins.
At high
temperatures the system is in the paramagnetic phase with free energy
\begin{equation}
\label{e2}
\ln Z=\alpha N\ln cosh(\beta)+N\ln 2
\end{equation} 
At critical temperature $T_c=1/\beta_c$ a phase transition occurs into 
the SG phase
\begin{equation}
\label{e3}
\alpha g(\tanh \beta_c)=1
\end{equation} 
where the function $g(x)$ is defined as 
\begin{equation}
\label{e4}
g(x)=\frac{1}{2}(1+x)\ln (1+x)+\frac{1}{2}(1-x)\ln(1-x)]
\end{equation}
Below this temperature 
we have 
\begin{equation}
\label{e5}
\ln Z/N=\alpha N\beta y(\ln2/\alpha)
\end{equation} 
where $y$ is inverse function to $g(x)$
\begin{equation}
\label{e6}
\frac{1}{2}(1+y)\ln (1+y)+\frac{1}{2}(1-y)\ln(1-y)]
=x
\end{equation}
\section{Dilute GREM}
Let us consider the diluted version of GREM, with infinite levels of the
hierarchy. Now we have an infinite chain of DREM-s. At the interval
$[v,v+dv]$ we have $Ns'(v)/\ln 2$ spins with $Nz d v$ couplings. Function
$s(v)$ is monotonic, $s(0)=0$,$s(1)=\ln 2$.
Similarly to the case of
dilute GREM at real T we found:
\begin{equation}
\label{e7} 
-\frac{\beta F}{N}=z(1-v_c(\beta)) \ln \cosh \beta
+(\ln 2-s(v_c)
+ z \beta\int_{0}^{v_2(\beta)}{\rm d}v_0 y(\frac{s'(v_0)}{z})
\end{equation}
where
$v_c(\beta)$
is defined from the equation
\begin{equation}
\label{e8}
\tanh(\beta)=y(s'(v_2)/z)
\end{equation}

For the case of Edwards-Anderson model placed on d-dimensional hypercubic
lattice
\begin{equation}
\label{e9}
z=d,\quad v=-\frac{U}{Nd}, \quad s(v)=\ln 2-\frac{S(-vdN)}{N}
\end{equation}
here $U$ is energy, and $S(U)$ is entropy as function of the energy for
corresponding ferromagnetic Ising model.
It is easy to derive from the definition of temperature 
\begin{equation}
\label{e10}
\frac{{\rm d} S}{{\rm d } U}=\frac{1}{\tau}\equiv \tilde\beta
\end{equation}
So there is a connection between $\tilde\beta$ and $v$. For a given
$\tilde\beta$ we find energy of corresponding ferromagnetic model, and 
then calculate 
\begin{equation}
\label{e11}
v=-U( \tilde\beta(v))/(Nd)
\end{equation}
We obtain for the free energy
\begin{equation}
\label{e12} 
-\frac{\beta F}{N}=z(1-v_c(\beta)) \ln \cosh \beta
(\ln 2-n(v_c(\beta)))
+ z \beta\int_{0}^{v_c(\beta)}{\rm d}v_0 y(\tilde\beta(v_0))
\end{equation}
\section{Complex temperatures}
Let us consider now the case of complex temperatures.
Now we have 3 phases for REM.\\
PM:
\begin{equation}
\label{e13}
\frac{\ln Z}{N}=\alpha  Re\ln cosh(\beta)+\ln 2
\end{equation} 
SG:
\begin{equation}
\label{e14}
\frac{\ln Z}{N}=\alpha \beta_1 y(1/\alpha)
\end{equation} 
LYF:
\begin{equation}
\label{e15}
\frac{\ln Z}{N}=\frac{\alpha}{2}  Re\ln cosh(2\beta_1)+\frac{\ln 2}{2}
\end{equation} 
 There are boundaries between SG- PM, SG - LYF ,PM - LYF.\\
For the PM-LYF we have a line 
\begin{equation}
\label{e16}
\sin^2(\beta_2)=\frac{2^{1/\alpha}-1-\tanh^2\beta_1}
{2^{1/\alpha}(1-\tanh^2\beta_1)}
\end{equation}
This line begins at $\beta_1=0$ and is contined till intersection with
SG-LYF
line. For it we have 
\begin{equation}
\label{e17}
\beta_1=\beta_0,\infty>\beta_2>\beta_{2c}
\end{equation}
where $\beta_{2c}$ is defined from the intersection with another line, and
$\beta_{1c}$ from the equation
\begin{equation}
\label{e18}
\alpha/2 \ln \cosh(2\beta_{1c})+\ln 2=\alpha \beta_{1c} y(\ln 2/\alpha)
\end{equation}
Then the third line PM-SG is defined from the equation
\begin{equation}
\label{e19}
\alpha Re\ln \cosh \beta+\ln 2=\beta_{1c}\alpha y(\ln 2/\alpha)
\end{equation}
Let us vary the parameter $\alpha$. We can construct some universal line
for the critical $\beta_{c,1},\beta_{c,2}$. If we define function 
\begin{equation}
\label{e20}
f(s,t)=\frac{\\ln(1+t)^2-\ln(1-t)^2[1-(1-t^2)s]}{\ln(1+t)/(1-t)}
\end{equation}
then for the critical $t=\tanh(\beta_1),s=\sin^2(\beta_2)$ we have an
equation
\begin{equation}
\label{e21}
\frac{1}{2}(1+f(s,t))\ln (1+f(s,t))+\frac{1}{2}(1-f(s,t))\ln(1-f(s,t))]
=\ln(1+t^2)/[1-(1-t^2)s]
\end{equation}

\section{DGREM at complex T}
Now we have 
\begin{equation}
\label{e22}
-\frac{\beta F}{N}=d(1-v_2)Re\ln \cosh \beta+(\ln2-s(v_2))+
d(v_2(\beta)-v_1(\beta))/2\ln \cosh 2\beta_1
\end{equation}
$$+[s(v_2)-s(v_1)]/(2)
+d\beta_1\int_{0}^{v_1}{\rm d}v_0
y({\tilde\beta(v_0)})$$
The values of $v_1,v_2$ are defined from the extremum condition.
Integrating by parts in the last term we get
\begin{equation}
\label{e23}
-\frac{\beta F}{N}=d(1-v_2(\beta))Re\ln \cosh \beta+(\ln2-s(v_2))+
d(v_2-v_1)/2\ln \cosh 2\beta_1
\end{equation}
$$+(s(v_2(\beta)-s(v_1(\beta)))/2
-d\beta_1 \int_{0}^{\tilde \beta_1}{\rm d}
\tilde \beta_0 \frac{2v_0(\tilde\beta_0)}{\ln\frac{1+y}{1-y}}
+d\tilde v_1\beta_1 y(\tilde\beta_1)$$
where $y$ as a function of $\tilde\beta_0$  is defined from the equation
\begin{equation}
\label{e24}
y=g(\frac{\ln2}{\tilde \beta_0})
\end{equation}
function $v_0(\tilde\beta)$ is defined from the equation
\begin{equation}
\label{e25}
v_0({\tilde \beta})
=-U(\tilde \beta)/(Nd)
\end{equation}
Here  $U(\tilde \beta_1)$ is the energy of ferromagnetic model
on the same lattice. The variables $v_1,v_2$ we can define from the saddle
point condition of free energy.\\
Let us collect all these results together.
We consider some point of hierarchy $v$. At inverse temperatures $\beta_1$
it could be in SG or PM phase. If it was in SG phase, then it stays there
while adding imaginary $\beta_2$.\\
When at real $\beta$ model was in PM phase there are three possible
scenarios of behavior: 1) stay in PM: 2) become
SG phase; 3)become LYF.
We can look under another point. What happens with our system, when we
vary $\alpha$ at fixed values of $\beta_1,\beta_2$?\\
If our point at complex $\beta$ space is under the line $\ref{e21}$, then
our system could be in either SG or PM phase. Above that line all 3 phases
are allowed for a different parts of hierarchy.\\
This line stays for any version of dilute GREM models. 
Thus it means some universality.
It will be interesting to find similiar 
universality classes in other mean field models.

\section{LYF zeros at the border of PM SG phases}
Let as first consider the case, when our system is under the line (21)
and  there is not LYF phase. We have an
expression 
\begin{equation}
\label{e26} 
\ln Z/N=d(1-v_c) Re\ln \cosh \beta
+\ln 2-s(v_c)
+ d \beta_1\int_{0}^{v_c}{\rm d}v_0 y({\tilde\beta(v_0)})
\end{equation}
where the value of $v_c$ is defined from the saddle point condition
\begin{equation}
\label{e27} 
d Re\ln \cosh \beta+s'(v_c)
=d \beta_1y(\frac{s'(v_c)}{z})
\end{equation}
For the laplacian we have an expression
\begin{equation}
\label{e28} 
(\frac{d^2f}{d\beta_1^2}+\frac{d^2f}{d\beta_2^2})=
\frac{\partial^2f}{\partial^2v_c}[(\frac{d v_c}{d\beta_1})^2+
(\frac{d v_c}{d\beta_2)^2}])+2\frac{\partial^2f}{\partial v_c\partial
\beta_1}(\frac{d v_c}{d\beta_1})+2\frac{\partial^2f}{\partial v_c\partial
\beta_2}(\frac{d v_c}{d\beta_2})
\end{equation}
To calculate density we need in expressions
\begin{equation}
\label{e29} 
s'(v)=d{\tilde\beta},
\quad y'=1/{\tilde\beta},\quad 
s''(v)=-\frac{d^2}{c({\tilde\beta})},
\quad y(\tilde{\beta_1})=\frac{Re \ln \cosh\beta+\tilde{\beta_1}}{\beta_1}
\end{equation}
here $c$ is specific heat of ferromagnetic phase. Eventually:
\begin{equation}
\label{e30} 
\pi\rho(\beta_1,\beta_2)=
\end{equation}
$$-\frac{d^2}{s''(-1+\beta_1y')}\{[(y-\tanh \beta_1-\frac{\tanh 
\beta_1\sin^2\beta_2}{\cosh
\beta_1^2(\sin^2\beta_2+\cos^2\beta_2\tanh^2\beta_1)}]^2
$$
$$+[\frac{\sin\beta_2\cos\beta_2}{\cosh
\beta_1^2(\sin^2\beta_2+\cos^2\beta_2\tanh^2\beta_1})]^2\}$$

\section{Three phases}
When our point in complex $(\beta_1,\beta_2)$ space is over the 
universality line,we have the follwoing expression for the laplacian:
\begin{equation}
\label{e31} 
(\frac{d^2f}{d\beta_1^2}+\frac{d^2f}{d\beta_2^2})=
f^{''}_{\beta_1\beta_1}+2f^{''}_{\beta_1v_1}v_{1,\beta_1}'+
f^{''}_{v_1v_1}(v_{1,\beta_1}')^2+2f^{''}_{\beta_1v_2}v_{2,\beta_1}'+
f^{''}_{v_2v_2}(v_{2,\beta_1}')^2
\end{equation}
$$+f^{''}_{\beta_2\beta_2}+2f^{''}_{\beta_2v_2}v_{2,\beta_2}'+
f^{''}_{v_2v_2}(v_{2,\beta_2}')^2$$
For the $v_1$ we have an equation
\begin{equation}
\label{e32} 
\frac{s'(v_1)}{2}+\frac{d}{2}\ln \cosh2\beta_1=\beta_1y(\frac{s'(v_1)}{d})d
\end{equation}
Having values $v_1,v_2$, wwe can define the values of corresponding 
$\tilde\beta_1,\tilde\beta_2$ by means of formula (25).
Equation for the $v_2$:
\begin{equation}
\label{e33} 
s'(v_2)+d\ln\cosh\beta_1+d/2\ln (\cos^2\beta_2+\tanh^2\beta_1\sin^2\beta_2)=
s'(v_2)/2+d/2\ln \cosh 2\beta_1
\end{equation}
Eventually we have for zeros density:
\begin{equation}
\label{e34} 
\pi\rho(\beta_1,\beta_2)=\frac{2z(v_2-v_1)}
{\cosh^2(2\beta_1)}-\frac{d^2(-\tanh2\beta_1+y(\tilde\beta_1))^2}{(\beta_1y'-1/2)s''(v_1)}
\end{equation}
$$+\frac{2d^2}{s''(v_2)}\{[\frac{(\sin\beta_2\cos\beta_2)}{[(\cosh\beta_1^2
(\cos\beta_2^2+\tanh\beta_1\sin\beta_2^2)}]^2$$
$$+[\tanh2\beta_1-\tanh\beta_1
-\frac{\sin^2\beta_2\sinh\beta_1}{\cosh\beta_1^3
(\cos\beta_2^2+\tanh\beta_1\sin\beta_2^2)}]^2\}$$

One can use formulas (30)  and (34) to calculate density of zeros, 
having the data of the ferromagnetic model at the same lattice.

\section{Conclusion.}
This paper devoted to the approximate method, which allows to investigate 
the zeroes
of the statistical sum
for Edwards-Anderson models in real physical dimensions.
The key point of the method is in the using the rich phase spectrum of 
Dilute Generalized Random Energy Model. We obtained a universality line 
in the phase diagram
of the model. It hardly controls the corresponding complex-temperature 
behavior.
The various notions that entered our discussion give good hope for the 
applicability of Random Energy-type models to realistic systems.

\begin{center} {\Large \bf Acknowledgements} \end{center}
D.B. Saakian is grateful to
Fundacion Andes grant c-13413/1 for financial support.  A. Crisanti, S.
Kobe and W. Janke are acknowledged for kind hospitality.

E.E. Vogel is grateful to Fondecyt 1990878.

\end{document}